\documentclass[aps,prl,superscriptaddress,showpacs,amsmath,amssymb,amsfonts,altaffillsymbol,twocolumn,nofootinbib]{revtex4-1}

\usepackage{graphicx}
\usepackage{dcolumn}

\graphicspath{{./figures/}}

\begin{document}
\title{
Yield in Amorphous Solids: The Ant in the Energy Landscape Labyrinth
}
\author{Ido Regev}
\email[E-mail: ]{regevid@bgu.ac.il}
\affiliation{Jacob Blaustein Institutes for Desert Research, Ben-Gurion University of the Negev,\\ Sede Boqer Campus 84990, Israel}

\author{Turab Lookman}
\affiliation{Theoretical Division, Los Alamos National Laboratory,\\ Los Alamos, New Mexico 87545, USA}

\date{\today}
\begin{abstract}
It has recently been shown that yield in amorphous solids under oscillatory shear is a dynamical transition from asymptotically periodic to asymptotically chaotic, diffusive dynamics. However, the type and universality class of this transition are still undecided. Here we show that the diffusive behavior of the vector of coordinates of the particles comprising an amorphous solid when subject to oscillatory shear, is analogous to that of a particle diffusing in a percolating  lattice, the so-called ``ant in the labyrinth'' problem, and that yield corresponds to a percolation transition in the lattice. We explain this as a transition in the connectivity of the energy landscape, which affects the phase-space regions accessible to the coordinate vector for a given maximal strain amplitude. This transition provides a natural explanation to the observed limit-cycles, periods larger than one and diverging time-scales at yield.
\end{abstract}
\maketitle

Plastic deformation in amorphous and granular materials occurs in many important applications in science and engineering, the most common of which are related to the mechanical properties of bulk metallic glasses, soft materials such as polymers, colloids and emulsions \cite{de1992soft} and to soil compaction \cite{richard2005slow} and consolidation in soil physics and civil engineering \cite{schofield1968critical}. Amorphous solids and granular materials are the result of a glass transition and jamming phenomena respectively \cite{debenedetti2001supercooled,liu1998nonlinear,berthier2011theoretical,boschan2016beyond,segall2016jamming}. The glass transition causes the system to become frozen in a metastable random configuration that is a local minimum of the potential energy. When these materials are subject to strain at a constant strain-rate, the stress typically increases but eventually settles into a steady-state with a constant value. The transition from elastic to plastic response is called ``yield'' \cite{schuh2007mechanical,bi2011jamming,bonn2015yield}. Recently, it has been discovered that under oscillatory shear, the yield point behaves as a non-equilibrium critical point separating a regime in which the system reaches a periodic limit-cycle, and a regime in which the system is always chaotic \cite{regev2013onset,regev2015reversibility,keim2013yielding,Granular1,nagamanasa2014experimental,kawasaki2015macroscopic,fiocco2013oscillatory,priezjev2016reversible,pfeifer2015order,leishangthem2016yielding,priezjev2016collective}. An implication of the periodicity below yield is that the system is not ergodic in that regime and for that reason starting from different initial conditions, the system reaches different steady-states as was shown by Fiocco et al. \cite{fiocco2013oscillatory}. Above yield the behavior is asymptotically diffusive \cite{fiocco2013oscillatory,berthier2011theoretical} with a diffusion coefficient that is zero at the transition and shows a power law dependence with the maximal strain amplitude. In previous work it has been suggested that the irreversibility transition is a phase transition into an absorbing state \cite{corte2008random,nagamanasa2014experimental}. Here we will show that the transition from periodic to diffusive behavior is actually a manifestation of an underlying percolation transition in the energy landscape/phase-space of the system, rather than a real-space percolation and that this can explain most of the phenomenology observed (limit-cycles, periods larger than one and ergodicity).

\begin{figure}[h]
  \begin{center}
    \includegraphics[width=\columnwidth]{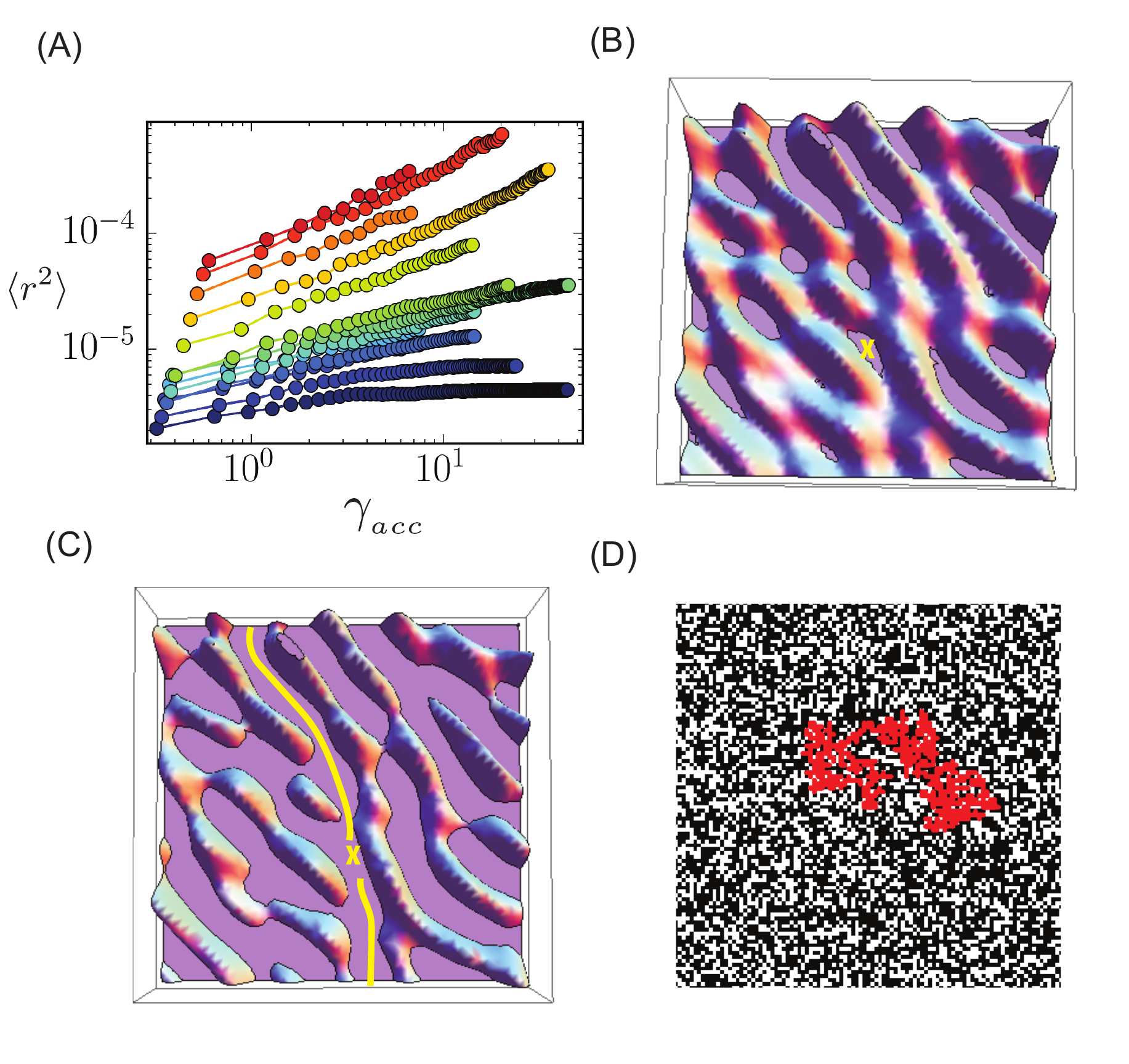}
    \caption{(color online) {\bf Percolation in the energy landscape} (A) $\langle {\bf r}^2\rangle$ from molecular dynamics as a function of accumulated strain $\gamma_{acc}$ for different maximal strain amplitudes $\gamma_{_{max}}=0.08,0.085,0.088,0.09,0.093,0.095,0.097,0.1,0.11,0.12,0.13,$\\$0.14,0.15$. Colors represent the magnitude of the maximal strain amplitude from cold to warm. (B) A two-dimensional caricature of the energy landscape before the percolation transition. Each point represents a configuration of the particles (in the simulations the surface $\mathcal{U}$ is embedded in a $2N$-dimensional coordinate space). For a given maximal strain amplitude, the regions accessible to the system can be considered as a ``sea'' limited by the geography of the landscape. If the system starts at the point marked in ``X'', it will not be able to reach the boundaries. (C) For a large enough maximal strain amplitude, the ``flooded'' regions connect to form a percolating basin, and starting from the marked ``X'' the system can reach the upper and lower boundaries. (D) An illustration of the path of a ``walker'' in a percolating cluster. Dark squares indicate sites that are occupied and the red points indicate sites visited by the diffusing walker. The walker can only diffuse from one dark (occupied) site to a neighbouring dark site.}
     \label{Fig1}
  \end{center}
\end{figure}

To study the transition, we simulated a system of $N$ ($N=16384$ in the simulation results shown) point particles in two dimensions interacting by a radially-symmetric attractive-repulsive potential. To avoid crystallization, we set half the particles to have a radius $1.4$ larger than the other half. To create an amorphous solid, we first simulated the system at a high temperature, in which the system is in a liquid phase, and then quenched the system to zero temperature using a minimization algorithm (FIRE\cite{FIRE}). We then deformed the system using the Athermal Quasi-static Shear (AQS) protocol, in which the dynamics comprise of minute shearing steps (shearing is performed by changing the boundary conditions using the Lees-Edwards scheme \cite{lees1972computer}) that are followed by relaxing the system to the next energy minima. In this way, we can increase the shear strain while keeping the system at effectively zero temperature. We performed cyclic shear by increasing the strain in AQS steps of $\delta\gamma=10^{-4}$ to a maximal strain amplitude $\gamma_{max}$ and then reduced the strain with the same small steps (now $-\delta\gamma$) applied in the negative direction until reaching a minimal strain amplitude $-\gamma_{max}$. We then reversed the straining direction again and increase the strain with the same $\delta\gamma$ steps to zero strain. This forms one cycle (a more detailed description of the simulations and potentials used can be found at Regev et al.\cite{regev2013onset}). All the simulation results in the paper were obtained by averaging over $30$ different amorphous solid realisations. To study the diffusive behavior of the system, we follow Fiocco et al.~\cite{fiocco2013oscillatory} and Kawasaki et al.~\cite{kawasaki2015macroscopic} and examine the Mean Square Displacement (MSD) of the particles:
\begin{equation}
\langle r(\gamma_{_{acc}})\rangle = \langle\sum_i^N|{\bf r}_i(\gamma_{_{acc}}) - {\bf r}_i(0)|^2\rangle\, ,
\end{equation} 
where $N=16384$ is the number of particles in the system. This provides a measure of how much each particle diffuses after an accumulated amount of strain $\gamma_{_{acc}}$ is applied \footnote{The accumulated strain is the sum of the absolute value of the number of strain steps $\delta\gamma$ performed since the start of the simulation}. The MSD shows the transition from transient, anomalous behavior below yield, to fully diffusive behavior above yield (Fig~\ref{Fig1}(A)) observed previously \cite{fiocco2013oscillatory,kawasaki2015macroscopic}. We have suggested before \cite{regev2015reversibility} that the transition from periodic to chaotic behavior is a result of a topological transition in the energy landscape. Here we will show that the reason for the diffusive behavior exhibited by these systems is a percolation transition below which the phase-space volume accessible to the system, starting from different initial conditions, is finite, and above which the accessible phase-space becomes infinite. A natural way in which this transition can be understood is in terms of the energy landscape. Due to the external forcing, a system that starts from a given particle configuration, will move to parts of the configuration/phase-space close to it (Fig~\ref{Fig1}(B)). As the external forcing on the system increases, energy barriers are diminished, which allows the system to explore larger and larger parts of the coordinate space. Eventually areas accessible from different initial conditions merge and an infinite connected cluster of available configurations emerges (Fig~\ref{Fig1}(C)). 

\begin{figure}[h]
  \begin{center}
    \includegraphics[width=\columnwidth]{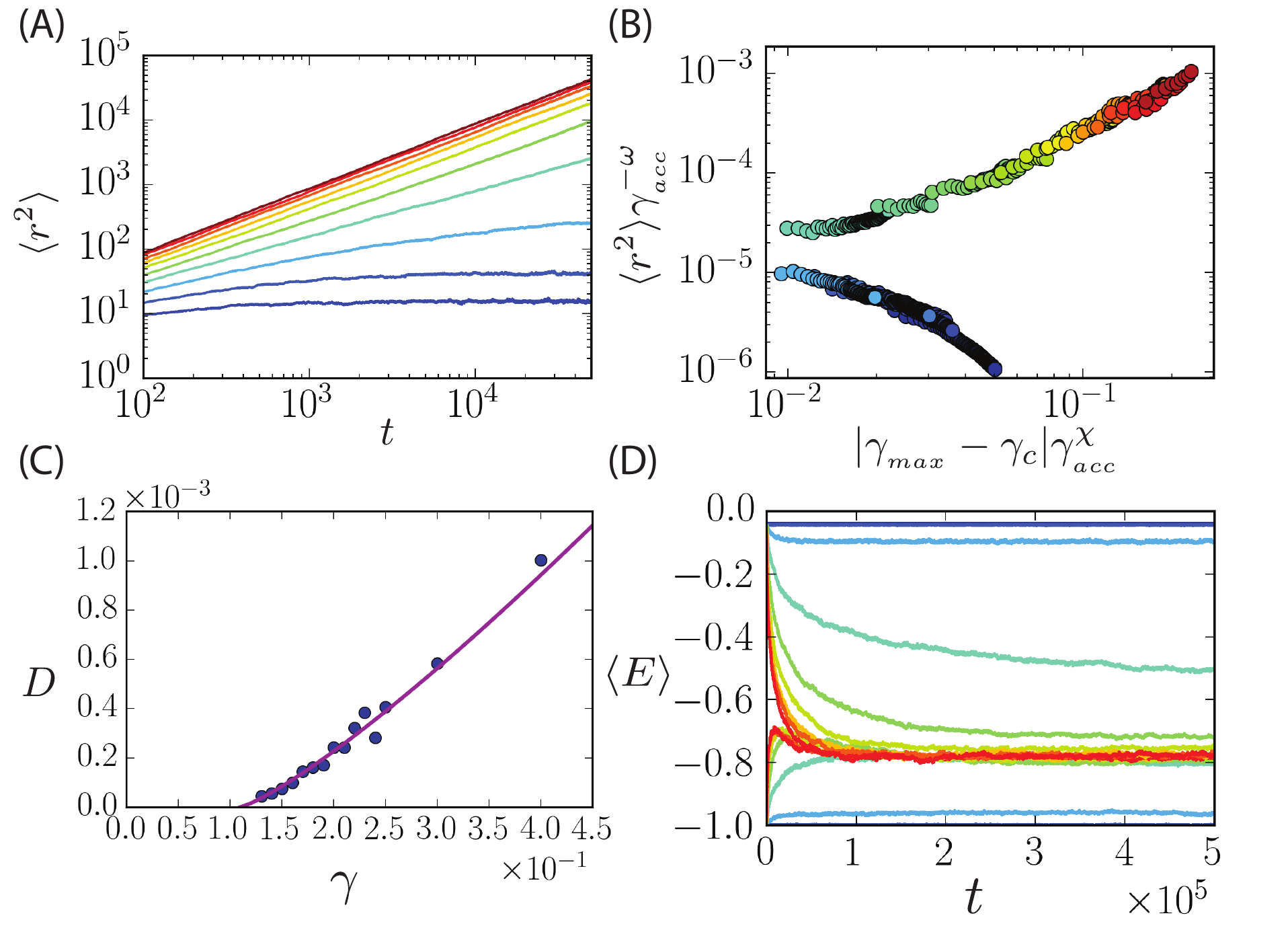}
    \caption{(color online) {\bf Diffusion}: (A)  $\langle {\bf r}^2\rangle$ as a function of time for a particle diffusing on a two dimensional percolating lattice. Colors indicate occupation probability $p$: from cold (blue) to worm (red), $p=0.45,0.5,0.55,0.6,0.65,0.7,0.75,0.8,0.85,0.9$. (B) Data collapse for $\gamma_{_{max}}=0.07,0.085,0.088,0.09,0.093,0.095$,\\$,0.097,0.1,0.11,0.12,0.13,0.14,0.15,0.16,0.17,0.18,0.19$,\\$,0.2,0.21,0.22,0.23,0.24,0.25$. Colors indicate the maximal strain amplitude $\gamma_{_{max}}$ - from cold (blue) to worm (red). (C) Diffusion coefficient from molecular dynamics for $\gamma_{_{max}}>\gamma_c$ fitted to a power law $D\sim(\gamma_{_{max}}-\gamma_c)^{\mu}$ where $\mu\approx 1.217$. (D) Energy as a function of (Monte Carlo) time for a particle diffusing on a two dimensional percolating lattice where each lattice site has a typical energy (see text). Colors indicate occupation probability $p$ - from cold (blue) to worm (red), $p=0.4,0.45,0.5,0.55,0.6,0.65,0.7,0.75,0.8$. For $p<p_c$ the steady-state potential energy depends on the initial conditions, while for $p>p_c$ the steady-state is always the same. This is a result of the system regaining ergodicity at $p_c$. The results are averages over 100 realizations.}
     \label{Fig2}
  \end{center}
\end{figure}

To show that the diffusive behavior of the system is indeed a result of an underlying percolation transition, we refer to the classical problem of a particle diffusing in a percolating network (Fig~\ref{Fig1}(D)) \cite{havlin1987diffusion,stauffer1994introduction}. This problem was first studied by De-Gennes who coined it ``the problem of the ant in the labyrinth'' \cite{de1976percolation} since it resembles the process in which a ``blind'' ant finds its way out of a labyrinth. The particle/ant starts on a  site in a lattice of sites that are either occupied or empty. At each time step (Monte Carlo step) it tries to move to one of $z$ neighbouring sites ($z$ depends on the lattice dimensionality) with equal probability. It can only diffuse into an occupied site. The neighbouring sites are occupied with probability $p$ which serves as the control parameter for the percolation transition. As $p$ is increased, larger and larger clusters of connected sites are formed. At $p=p_c$ these clusters connect to form an infinite spanning cluster which connects the entire system. Due to the fractal nature of these clusters, the system exhibits scaling behavior near the critical point; specifically, there is a diverging correlation length $\xi \sim |p-p_c|^{-\nu}$ which is related to the size of the clusters. The probability of a random site to be in a spanning cluster scales as:
\begin{equation}
P_{\infty}(p) \sim (p-p_c)^{\beta}\,,
\end{equation} 
and the conductivity of the material for $p>p_c$ scales as: 
\begin{equation}
\sigma(p) \sim (p-p_c)^{\mu}\,,
\end{equation}
($\sigma=0$ for $p<p_c$) \cite{stauffer1994introduction,gefen1983anomalous,havlin1987diffusion,havlin1983scaling}. If we define an asymptotic diffusion coefficient:
\begin{equation}
D(p) = \lim_{t\rightarrow\infty}\langle{\bf r}^2(t)\rangle/t\, ,
\end{equation}
(where $\langle{\bf r}^2\rangle$ is the average of the mean square displacement ${\bf r}^2(t) = \sum_i |r_i(t) - r_i(0)|^2$) it will have a different value depending on $p$. For $p<p_c$ the system will be in one of many disconnected clusters so that asymptotically $r^2(t)$ will reach a finite value:
\begin{equation}
\langle {\bf r}^2(t=\infty)\rangle \sim (p_c - p)^{\beta-2\nu}\, .
\end{equation} 
For $p>p_c$ there is an infinite cluster spanning the system and therefore there will be a non zero diffusion coefficient at asymptotic times (Fig~\ref{Fig2}(A)). 
The asymptotic diffusion coefficient will depend on $p$ since even above the transition there are still quite a few finite clusters and a particle starting on one of the finite clusters will not contribute to the asymptotic diffusion. For this reason, together with Einstein's relation, we expect that the diffusion coefficient will have the following scaling behavior for $p>p_c$ \cite{havlin1987diffusion}:
\begin{equation}
D(p) \sim \sigma(p) \sim (p - p_c)^{\mu}\, .
\end{equation}

To demonstrate that the diffusion in the sheared amorphous solid corresponds to a diffusion on a percolating network, we performed a scaled data collapse to the scaling relation:
\begin{equation}
\langle {\bf r}^2\rangle \gamma^{-\omega} = \mathcal{F}_{\pm}\left(|\gamma_{_{max}}-\gamma_c|\gamma^{\chi}\right)\, ,
\end{equation}
where $\mathcal{F}_+(z)$ and $\mathcal{F}_-(z)$ are the two branches of the crossover scaling function for below and above the transition and $\gamma_c$ is the maximal strain amplitude at yield. This scaling relation was shown to describe the MSD close to the percolation transition \cite{havlin1987diffusion,stauffer1994introduction}. Using this function we obtain a data collapse into the two branches of the scaling function (see Fig~\ref{Fig2}(B)). Using the data collapse we also extracted the exponents $\omega$ and $\chi$ that are functions of the percolation exponents. We find:
\begin{equation}
\omega = \frac{2\nu - \beta}{2\nu + \mu - \beta} \approx 0.6\,,
\label{omega}
\end{equation} 
and 
\begin{equation}
\chi = \frac{1}{2\nu + \mu - \beta} \approx 0.22\,,
\label{chi}
\end{equation} 
so that the MSD can be described using the expression:
\begin{equation}
\langle {\bf r}^2\rangle  = \gamma^{0.6}\mathcal{F}_{\pm}\left(|\gamma_{_{max}}-\gamma_c|\gamma^{0.22}\right)\,.
\label{scaling}
\end{equation}
To estimate the critical exponent $\mu$, we calculated the diffusion coefficient as a function of $\gamma_{_{max}}$ and fitted it to a power-law $D\sim (\gamma_{_{max}}-\gamma_c)^{\mu}$ (fig~\ref{Fig2}(C)), which gives $\mu\approx 1.217$. A conductivity exponent $\mu>1$ is typical of percolating systems \cite{kirkpatrick1973percolation} but is very different from the values $\mu\approx 0.54-0.61$ obtained by Fiocco et al. \cite{fiocco2013oscillatory} for simulations in three dimensions. A value of $\mu>1$ would appear to also be a reasonable fit to the data in \cite{fiocco2013oscillatory}. Currently, we are not able to find $\mu$ and $\nu$ from Eq.~\ref{omega},\ref{chi} since the equations are linearly dependent. This should be addressed in future research using other measures.\\

{\bf Discussion}
{\it Explanation for the observation of limit cycles}: The percolation transition picture provides a natural way to explain the appearance of limit cycles, the diverging time to reach a limit cycle at the transition and the appearance of periods larger than one. The reason for the observation of limit cycles is that below yield the accessible phase-space\footnote{For athermal deformation phase-space is the same as the coordinate space.} volume is finite and thus the phase-space trajectory of the system diffuses in a confined space with reflecting boundaries. The finiteness of the available phase-space guarantees that the phase-space trajectory will self-intersect after some finite time. Since the equations are deterministic and the randomness comes only from the complexity of the energy landscape, once the phase-space trajectory self-intersects, the system will repeat the same trajectory forever (it will enter a limit-cycle). For a small cluster size, the phase-space trajectory will self intersect after a short time. As we increase the maximal strain amplitude (which is equivalent to increasing the occupation probability $p$), the clusters of accessible phase-space become larger and larger and thus the coordinate vector can wander on the energy landscape for a longer time before the trajectory self-intersects. Furthermore, as the phase-space trajectory becomes larger, there is a larger probability that the system self-intersects after more than one shearing cycle, resulting in a period larger than one.

{\it Explanation of the ergodic properties of the system}: To test the idea that the ergodic properties of the system are described by diffusion on a percolating network model, we performed Monte Carlo simulations of a particle diffusing on a percolating lattice where each lattice site has an energy assigned. The energy function is chosen to be a half-Gaussian which is a function of only the $x$ direction: 
\begin{equation}
E(x,y) = -e^{-x^2/\sigma^2}\,, 
\end{equation} 
but other options may apply (one objective of future research will be to identify the right function). For $p<p_c$ starting from two different initial conditions, one with a high energy and one with a low energy, the steady-states of different initial conditions are different (blue curves in Fig~\ref{Fig2}(D)). However, for $p>p_c$ the steady-states are the same (red curves) which indicates that the system regains some form of ergodicity. This is very similar to the behavior observed in \cite{fiocco2013oscillatory} for the potential energy as a function of accumulated strain starting from two different initial quenches - a quench from a high temperature and a quench from a low temperature. For small maximal strain amplitudes, the average potential energy at the steady-state is different and depends on the initial state, but for maximal strain amplitudes $\gamma_{max}>\gamma_c$, the potential energy at the steady state does not depend on the initial quench (see Fig 1 of \cite{fiocco2013oscillatory}). An immediate implication of the percolation transition description is that in this model, the typical time-scale to reach the steady-state potential energy grows as the typical time-scale for the ``ant'' to reach the maximal cluster size. This is typically $t \sim (p_c-p)^{2\nu + \mu - \beta}$ for the ant in the labyrinth and we expect it to be $\gamma_{_{acc}} \sim (\gamma_c - \gamma_{_{max}})^{2\nu + \mu - \beta}$ for amorphous solids. A more accurate description is possible using the scaling relation Eq.~\ref{scaling}. One feature which was observed in amorphous solids \cite{fiocco2013oscillatory} and is not reproducible by the standard percolation-diffusion model is that the average potential energy keeps changing even for maximal strain amplitudes larger than the critical point. We believe that this is a result of correlations between the occupation probabilities of different sites (minima of the energy landscape) and we will address this in future work.

{\it Avalanches and fractal energy landscapes}: It has recently been suggested that the energy landscape of glassy materials has a fractal structure and that this fractal structure gives rise to avalanche behavior \cite{hwang2016understanding,charbonneau2014fractal,franz2016mean}. In previous work we have found that under periodic shear, avalanche sizes diverge at yield \cite{regev2015reversibility}. A possible connection between the two phenomena is that at yield, there are regions in which there are no energy barriers, (see Fig\ref{Fig1}(B,C)) and for this reason the susceptibility of the system to small perturbations becomes very large \cite{regev2015reversibility}. It will be interesting to study this connection in more detail.

{\it Connection to the glass transition}: It has been suggested that the glass transition is accompanied by a geometrical transition in the energy landscape. In this picture, above a certain temperature, the system spends most of the time around saddles of the energy landscape, while above the transition it spends most of the time near minima of the energy landscape \cite{grigera2002geometric}. While the glass transition is a thermal phenomenon, and we have studied an athermal phenomena, the picture in the energy landscape appears to be related and it will be interesting to explore the connections.\\

{\bf Acknowledgements} I.R. would like to thank Yair Shokef and Golan Bel for useful discussions. 

\bibliography{percolation}

\end{document}